\documentclass{JHEP3}
\usepackage{graphicx,amsmath,amssymb}
\usepackage{epsfig,multicol}
\usepackage{epsf}
\usepackage{epstopdf}
\input{epsf.sty}

\usepackage{graphicx}
\usepackage{amssymb}
\usepackage{amsmath}

\def\half{{1\over 2}}
\def\({\left(}
\def\){\right)}
\def\[{\left[}
\def\]{\right]}

\def\e{\begin{equation}}
\def\q{\end{equation}}
\def\m{\begin{eqnarray}}
\def\n{\end{eqnarray}}
\providecommand{\abs}[1]{\lvert#1\rvert}

\title{New features in curvaton model}
\author{Pravabati Chingangbam $^{1}$ \footnote{prava@kias.re.kr} \ and
  Qing-Guo Huang $^2$ \footnote{huangqg@itp.ac.cn}
\\ \small{\em $^1$ Astrophysical Research Center for the Structure and
  Evolution of
the Cosmos, Sejong University, 98 Gunja Dong, Gwangjin gu, Seoul
  143747, South Korea }

\\\small{\em $^2$
Key Laboratory of Frontiers in Theoretical Physics,
Institute of Theoretical Physics, Chinese Academy
of Sciences, Beijing 100190, China}
 }

\abstract{
We demonstrate novel features in the behavior of the second
  and third order non-linearity parameters of the curvature perturbation,
  namely, $f_{NL}$ and $g_{NL}$,
  arising from non-linear motion of curvaton field. We investigate two
  classes of potentials for the curvaton - the first has tiny
  oscillations super-imposed upon the quadratic potential. The second
  is
  characterized by a single `feature' separating two quadratic
  regimes with different mass scales. The feature may either be a bump
  or a flattening of the potential. In the case of the oscillatory
  potential we find that as the width and height of superimposed
  oscillations increase, both $f_{NL}$ and $g_{NL}$ deviate strongly
  from their expected values from a quadratic
  potential.  $f_{NL}$ changes sign from positive to negative as the
  oscillations in the potential become more prominent. Hence, this
  model can be severely constrained by convincing evidence from
  observations that $f_{NL}$ is positive. $g_{NL}$, on the other hand,
  acquires very large negative values. For the 
  the single feature potential, we find that
  $f_{NL}$ and $g_{NL}$ exhibit oscillatory behavior as a function of
  the parameter that controls the feature.
}


\begin{document}
%%%%%%%%%%%%%%%%%%%%%%%%%%%%%%%%%%%%%%%%%%%%%%%%%%%%%%%%%%%%%%%%%%%%%%%%%%%%%%

\section{Introduction}

The inflationary paradigm \cite{Guth:1980zm} has become an important
ingredient of modern cosmology. Inflation provides a natural
explanation for the production of
the first density perturbations in the early universe which
seeded the formation of the large scale structure (LSS) in
the distribution of galaxies and the temperature
anisotropies in the cosmic microwave background radiation
(CMBR) \cite{Guth:1982ec}. However, the precise details of the
mechanism for generating the primordial curvature perturbation is not
fully established. The standard mechanism is via the quantum
fluctuations of the inflaton field. An alternative scenario which
frees the inflaton from the job of generating perturbations, besides
giving rise to inflation, is the curvaton
scenario~\cite{Enqvist:2001zp,Lyth:2001nq,Moroi:2001ct}. The
curvaton is assumed to be a light scalar field which begins evolving
at the end of inflation. Its energy density is assumed to be
subdominant during inflation, but it can share a significant part of
the total energy in the universe before its decay.  The entropy
perturbations caused by the curvaton field finally get converted into
adiabatic perturbations.

A large number of light scalar fields are expected to be present in
any fundamental theory that goes beyond the standard model of particle
physics. During the inflationary era these fields would
have had the same amplitude of quantum fluctuations.
It is plausible that at least some of them played important roles in
the early universe, for example, as the curvaton field. An important
distinguishing property of the
curvaton scenario as the generating mechanism for primordial
perturbations, from the standard single slow-rolling field
picture, is the possibility for the
primordial perturbations to have large deviations from Gaussian
distribution. This property becomes very attractive
in the light of the recent result
from WMAP which suggests that primordial non-Gaussianity
may be large \cite{Komatsu:2010fb}.
The non-Gaussianity generated in the curvaton scenario must have a
local shape because it is generated on superhorizon
scales. Then the curvature perturbation can be expanded at the
same spatial point to non-linear orders, as,
\m
\zeta({\bf x})=\zeta_g({\bf x})+{3\over 5}f_{NL}\(\zeta_g^2({\bf
  x})-\langle  \zeta_g^2\rangle \)+{9\over 25} g_{NL}\(\zeta_g^3({\bf
  x})-3\langle
\zeta_g^2\rangle \zeta_g\)+... \ ,
\label{zetafg}
\n
where $f_{NL}$ and $g_{NL}$ are the so-called non-Gaussianity
parameters. The WMAP 7yr  result
implies a constraint on the size of local form bispectrum as
$f_{NL}=32\pm21$ at $1-\sigma$ level. The limits on $g_{NL}$
are $-3\times 10^{-5} < g_{NL} < 8\times 10^{-5}$ from
LSS~\cite{Desjacques:2009jb} and similar limits are obtained from CMB
data from WMAP 5yr data~\cite{Smidt:2010sv} as well.  A convincing
detection of the local form non-Gaussianity will rule out
all single-field inflation in a model-independent way.

In the simplest case the curvaton potential is assumed to
have a quadratic form and the typical size of the bispectrum is
bounded by the tensor-scalar ratio, as, $f_{NL}<10^3
r^{1/4}$, \cite{Huang:2008ze}.
Since the curvaton field
evolves linearly in this case, the size of the trispectrum is
linearly related to that of the bispectrum, as,
\e
g_{NL}\simeq -{10\over 3}f_{NL}.
\label{eq:gnlfnl}
\q
There is, however, no reason for the above relation to hold in general
from the viewpoint of fundamental theory.
Apart from the quadratic potential, all the curvaton models that have
been considered so far in the literature focus on potentials which
deviate from the quadratic form at large field values but tend to the
quadratic form at small field values. The predictions of such models,
particularly the level of non-Gaussianity, are then compared with
those from the quadratic potential so as to understand their
distinguishing features. Clearly, the distinction becomes more
prominent as the initial curvaton field value becomes larger and
larger. A distinct signature of departure of the curvaton potential
from quadratic form is a breakdown of the relation~(\ref{eq:gnlfnl}).
If the curvaton self-interaction term becomes dominant, giving rise to
higher order corrections in the curvaton potential, the order of
magnitude of $g_{NL}$ can be ${\cal O}(f_{NL}^2)$
\cite{Huang:2008zj,Enqvist:2009zf,Enqvist:2009eq,Enqvist:2009ww}.
The predictions of curvaton model with nearly quadratic
potential are investigated in
\cite{Enqvist:2005pg,Sasaki:2006kq,Enqvist:2008gk,Huang:2008bg},
where the non-linear evolution of curvaton after inflation
but prior to its oscillation is taken into account. Another
promising curvaton candidate is the pseudo-Nambu-Goldstone
boson -- axion, whose potential significantly deviates from
quadratic form around the top of its potential. A numerical
analysis of the axion-type curvaton model is discussed in
\cite{Kawasaki:2008mc,Chingangbam:2009xi}. From
the viewpoint of fundamental theory, one can generically expect
multi curvatons models and such a model is investigated in
\cite{Assadullahi:2007uw,Huang:2008rj}. While the discussion this far
has ignored scale dependence of the bispectrum and the trispectrum, it
is possible that in the future such scale dependences may become
accessible to experimental observation and hence important
\cite{Byrnes:2009pe}. Other papers of relevance are
\cite{Kawakami:2009iu,Matsuda:2009kp,Takahashi:2009cx,Nakayama:2009ce,
Kawasaki:2009hp,Cai:2010rt}.

We investigate two new curvaton models different from the ones
described above. The first is a potential which has tiny
oscillations superimposed on the quadratic form. The resulting effect
on the curvaton evolution is that it experiences the small bumps of
the oscillations in the potentials during the stage of it evolution
when it undergoes oscillations about the minimum of the potential. As
a consequence, the curvaton evolution during this stage
is non-linear (the curvaton equation of motion is not that of a damped
simple harmonic oscillator), making it significantly different from the
curvaton oscillation about the minimum of a quadratic potential.
Our goal is to calculate the non-linear curvature perturbation up to cubic
order and obtain the predictions for non-Gaussianity from such a model.
We find very interesting new implications for the non-linearity
parameters $f_{NL}$ and $g_{NL}$ arising in this model. First,
$f_{NL}$ is no longer restricted to have positive values. Depending on the
amplitude and the frequency of the superimposed oscillations on the
potential, it can take a wide range of both positive
and negative, with a switch of sign from positive to negative.
$g_{NL}$ , on the other hand, remains negative and can take large
negative values. The sign switch of $f_{NL}$ brings up the possibility
that the most important contribution to
primordial non-Gaussianity could come from the $g_{NL}$ term, with
$f_{NL}$ being negligibly small. 

The second model we discuss is a class of potentials characterized by
a single {\em feature}
separating two quadratic regimes with different mass scales. The
feature depends on a single parameter and depending on the sign of the
parameter, it can be either a single bump or a flattening of the slope
of the quadratic potential at some characteristic scale. We find that
the effect of the feature on $f_{NL}$ and $g_{NL}$ is rather dramatic,
causing them to oscillate with increasing amplitude as the strength of
the feature increases.

This paper is organized as follows: in section 2, we briefly summarize
the method for computation of the non-linear curvature perturbation
using the $\delta N$ formalism and the curvaton equation of motion. In
section 3, we describe the specific forms of the curvaton potentials
we are considering here and display our results for the non-linear
corrections to the curvature perturbations. In section 3.1, we discuss
the case of the washboard potential,
while in section 3.2, we discuss the single feature potential.
We end with a summary of our results and
comments in section 4. A brief description of the curvaton with
quadratic potential is given in the appendix to highlight the
differences from our study and novelty of our results.

%%%%%%%%%%%%%%%%%%%%%%%%%%%%%%%%%%%%%%%%%%%%%%%%%%%%%%%%%%%%%%%%%%%%

\section{The non-linear curvature perturbation}

On sufficiently large scales, the curvature perturbation on
the uniform density slicing can be calculated by using the
so-called $\delta N$ formalism
\cite{Starobinsky:1986fxa,Sasaki:1995aw,Sasaki:1998ug,Lyth:2004gb,Lyth:2005fi}.
Starting from any initial flat slice at time $t_{ini}$, on
the uniform density slicing, the curvature perturbation is
\e
\zeta(t,{\bf x})=\delta N\equiv N(t,{\bf x})-N_0(t),
\q
where $N(t,{\bf x})=\ln a(t,{\bf x})/a(t_{ini})$ describes
the local expansion of our universe, and $N_0(t)=\ln
a(t)/a(t_{ini})$ is the unperturbed amount of expansion.
In curvaton model, the difference between local expansion
and the unperturbed expansion is caused by the quantum
fluctuations of curvaton field during inflation. Therefore
\e
\zeta=N_{,\sigma}\delta \sigma+\half N_{,\sigma\sigma}{\delta
\sigma}^2+{1\over 6} N_{,\sigma\sigma\sigma} {\delta \sigma}^3+... \ ,
\q
where $N_{,\sigma}=dN/d\sigma$,
$N_{,\sigma\sigma}=d^2N/d\sigma^2$ and
$N_{,\sigma\sigma\sigma}=d^3N/d\sigma^3$. Considering
$\delta\sigma={H_*}/2\pi$, the amplitude of the power
spectrum generated by curvaton field is
\e
P_{\zeta_\sigma}=N_{,\sigma}^2 \({H_*\over 2\pi} \)^2,
\q
and the non-Gaussianity parameters are given by
\m
f_{NL}&=& {5\over 6}{N_{,\sigma\sigma}\over N_{,\sigma}^2},\\
g_{NL}&=& {25\over 54} {N_{,\sigma\sigma\sigma}\over N_{,\sigma}^3},
\n
where $H_*$ is the Hubble parameter during inflation.
On the other hand, the amplitude of the tensor perturbation
only depends on the inflation scale, namely
\e
P_T={H_*^2/M_p^2 \over \pi^2/2}.
\q
Thus the tensor-scalar ratio $r$ is given by
\e
r\equiv P_T/P_{\zeta_\sigma}={8\over N_{,\sigma}^2 M_p^2}.
\q
Here we consider the simplest version of curvaton scenario
where the quantum fluctuations of convaton field contribute
the total curvature perturbation.

After inflation, the equations of motion are
\m
H^2&=&{1\over 3M_p^2}(\rho_r+\rho_\sigma), \\
\dot \rho_r&+&4H\rho_r=0, \\
\rho_\sigma&=&\half {\dot \sigma}^2+V(\sigma), \\
\ddot \sigma&+&3H\dot \sigma+{dV(\sigma)\over d\sigma}=0,
\n
where $\rho_r$ and $\rho_\sigma$ are the energy densities
of radiation and curvaton respectively, and $V(\sigma)$ is
curvaton potential. In order to numerically solve the above
differential equations, we define the reduced curvaton
field $\tilde \sigma$ and reduced curvaton potential
$V(\tilde \sigma)$ as follows
\m
{\tilde \sigma}&=& \sigma/\sigma_*,\\
V(\tilde \sigma)&=& {V(\sigma)\over m^2\sigma_*^2},
\n
where $\sigma_*$ is the vacuum expectation value (VEV) of
curvaton field in the inflationary era.
Now the equations of motion can be simplified to be
\m
N'&=&\[\alpha e^{-4N}+{\sigma_*^2\over 3M_p^2}\(\half {\tilde
  \sigma}'^2 +V({\tilde \sigma})\)\]^\half, \label{eqn:nprime}\\
{\tilde \sigma}''&+&3N'{\tilde \sigma}'+{dV(\tilde \sigma)\over
  d{\tilde  \sigma}}=0, \label{eqn:tsigma}
\n
where $N(x)=\ln a(t)$, $\alpha={\rho_{r,ini}\over
3M_p^2m^2}={H_{ini}^2/m^2}$, and the prime denotes the
derivative with respect to dimensionless time coordinate
$x\equiv mt$, and the Hubble parameter becomes \e H=mN'. \q The
solution for the subdominant curvaton with quadratic
potential is analytically discussed in the appendix.

The scale factor can be rescaled to satisfy $a(t_{ini})=1$,
or equivalently $N(t_{ini})=0$. For numerical calculation,
we also need to input the value of $\alpha$. If the vacuum
energy of inflaton suddenly decays into radiation, a
reasonable choice is $\alpha=H_{inf}^2/m^2$ which is much
larger than one. However we don't know its value exactly.
But as long as $\alpha$ is large enough it does not affect
our numerical result because the curvaton field almost does
not move when the Hubble parameter is much larger than its
mass. For example, it is reasonable to assume that the
Hubble parameter at the inflationary era is one order of
magnitude larger than the curvaton mass at least and then
we set $\alpha=10^2$ in this paper.

%%%%%%%%%%%%%%%%%%%%%%%%%%%%%%%%%%%%%%%%%%%%%%%%%%%%%%%%%%%%%%
\section{The models}

In this section we consider two new curvaton models which
have some small features around the exactly quadratic
form of the curvaton potential. We can expect that these features will
introduce non-linear effects to the oscillating curvaton field and
consequently affect the non-Gaussianity parameters. Our aim is to
calculate the precise effects.
Note that these effects are different from what was considered in
\cite{Enqvist:2005pg,Sasaki:2006kq,Enqvist:2008gk,Huang:2008bg}
where the non-linear evolution of curvaton after inflation
but prior to its oscillation was considered.
Since the non-linear nature of the curvaton motion makes analytic
solutions extremely difficult to obtain, we rely on numerical methods
to get our results. We solve the Eqs.~(\ref{eqn:nprime}) and
(\ref{eqn:tsigma}) as a coupled set of differential equations for each
potential under consideration.

%%%%%%%%%%%%%%%%%%%%%%%%%%%%%%%%%%%%%%%%%%%%%%%%%%%%%%%%%%%%%%%%
\subsection{Washboard curvaton model}

Let us consider a curvaton potential which has tiny
oscillations superimposed on the exactly quadratic form. We call
it the {\em washboard model} and it takes the following explicit form
\e V(\sigma)=\half
m^2 \sigma^2+ V_0 \(1-\cos ({\sigma\over F})\), \q
where $V_0\ll V_*=\half m^2\sigma_*^2$. The reduced potential of
$\tilde \sigma$ is
\e
V(\tilde \sigma)=\half {\tilde \sigma}^2+\epsilon \(1-\cos({\tilde
  \sigma/\delta})\),
\label{eqn:pot_wb}
\q
where
\e
\epsilon={V_0\over m^2\sigma_*^2}, \quad
\delta={F\over \sigma_*}.
\q
\begin{figure}[h]
\begin{center}
\includegraphics[height=6.5cm,width=6.cm]{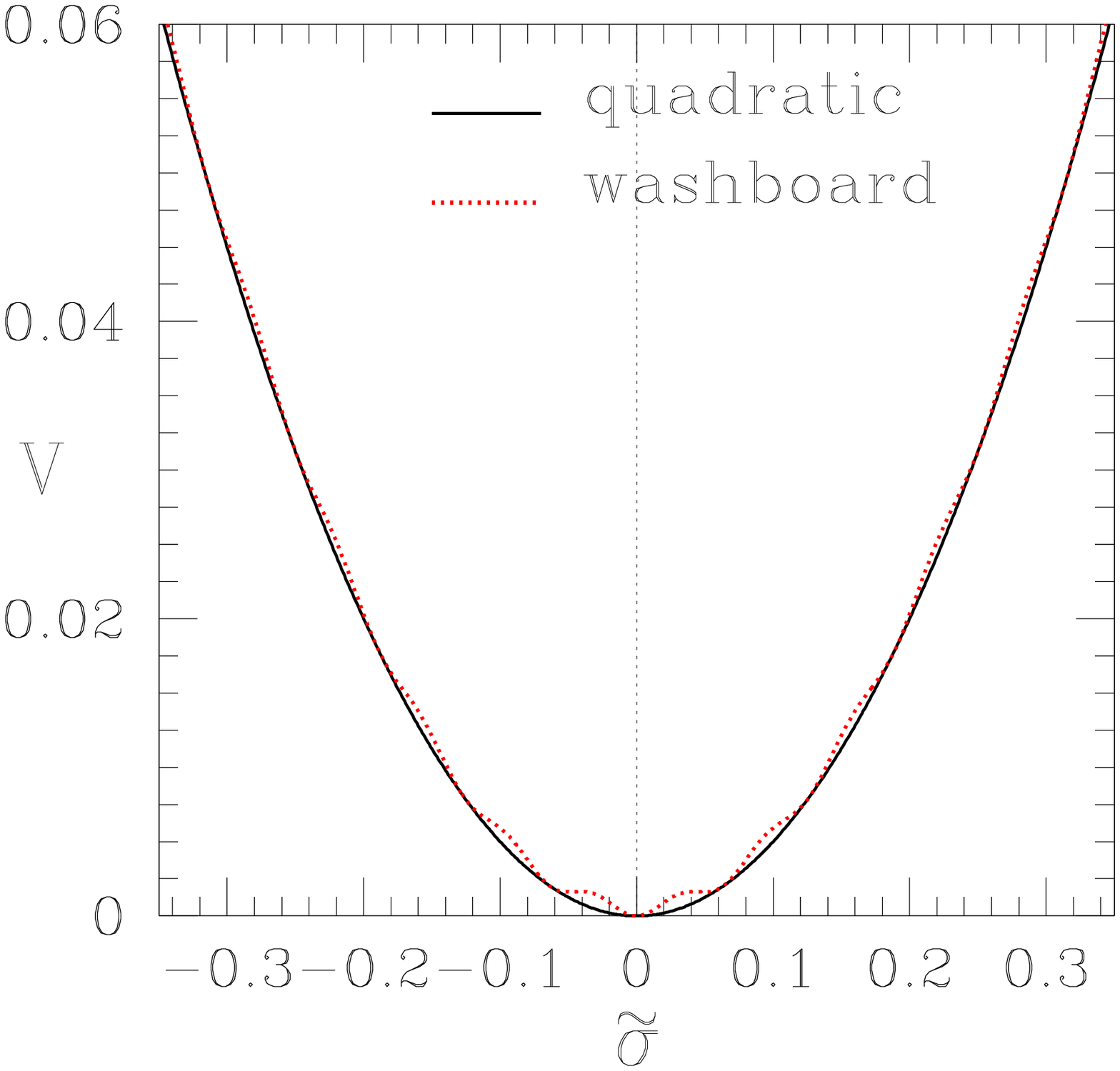}
\includegraphics[height=6.5cm,width=9.cm]{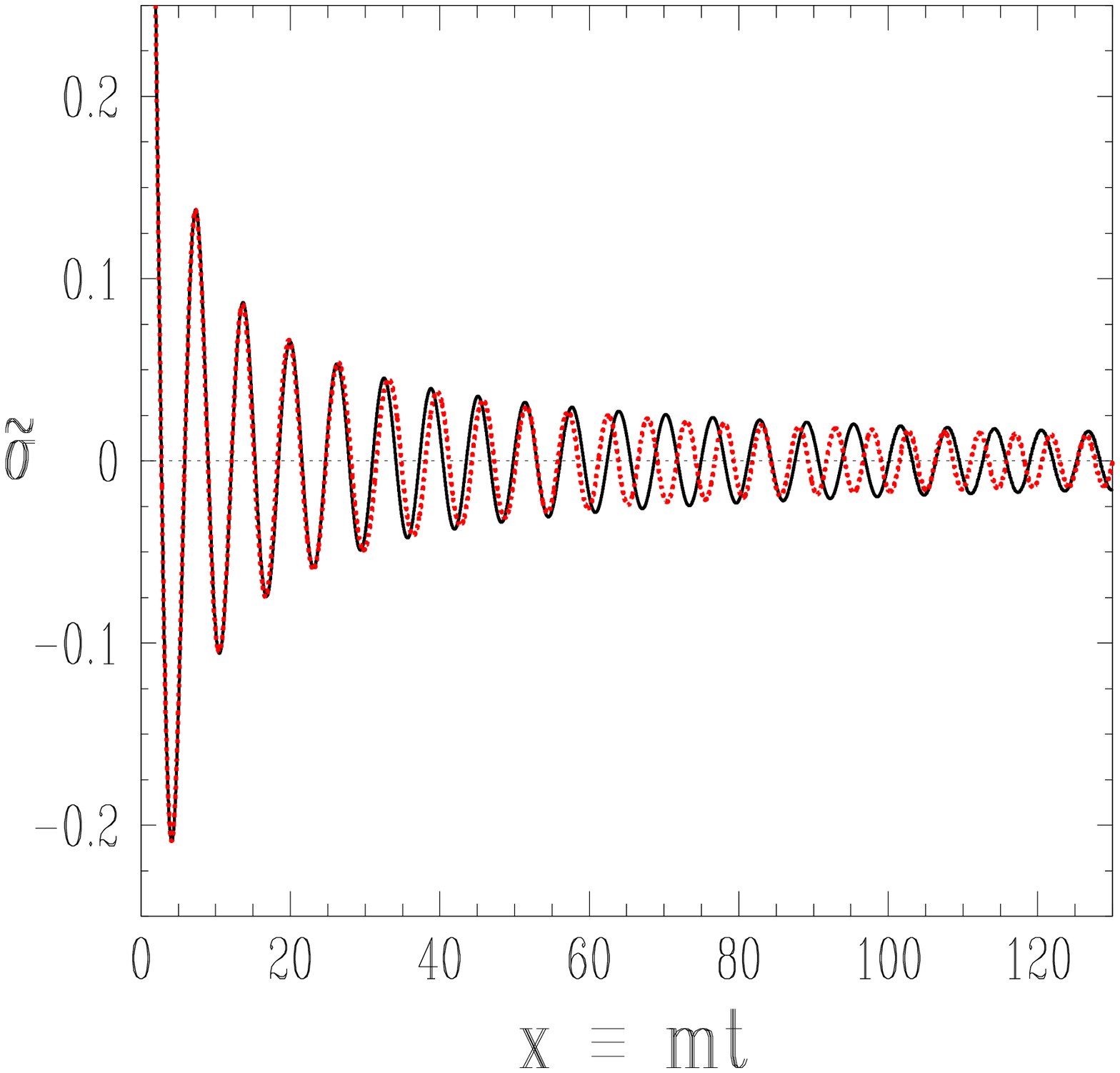}
\end{center}
\caption{The washboard curvaton potential given by
  Eq.~(3.2) is shown on the left panel for visual comparision with the
  corresponding quadratic one. The parameter values are
  $\epsilon=5\times 10^{-4}$ and $\delta=10^{-2}$. We have chosen
  a large value of $\epsilon$ in order to make the oscillations
  clearly visible. The right panel shows the curvaton
  oscillations about the potential minimum for the quadratic and
  washboard cases, with the same initial field value given by
  $\sigma_*/M_p=0.1$. The parameter values for this plot are
  $\epsilon=10^{-4}$ and $\delta=10^{-2}$}
\label{fig:pwb}
\end{figure}
Here $\epsilon$ measures the size of the correction and
$\delta$ characterizes the period of oscillation of the correction
term in the washboard potential.
The reduced potential is shown in the left panel of Fig. \ref{fig:pwb}
for easy visualization.
When ${\tilde \sigma}\gg \sqrt{\epsilon}$, the potential is
almost quadratic. If ${\tilde \sigma}\ll \delta$, then $V(\tilde
\sigma)\simeq \half (1+{\epsilon/ \delta^2}) {\tilde
\sigma}^2$. Then the curvaton potential is roughly quadratic
as well, but has a deformed mass.

The dynamics of curvaton field after inflation is governed
by \e {\tilde \sigma}''+{3\over 2x}{\tilde \sigma}'+{\tilde
\sigma}+ {\epsilon \over \delta}\sin\({\tilde \sigma/
\delta} \)=0. \q Even though the correction to the
potential is small, the dynamics of curvaton field becomes
significantly non-linear if the period of the correction
term is small enough. Here we consider the case in which
the dynamics of curvaton is dominated by the mass term in
the beginning, which implies $\epsilon/\delta <1$. Once the
amplitude of the curvaton oscillation drops below
$\epsilon/\delta$, the curvaton evolves non-linearly.
On the right panel of Fig.~\ref{fig:pwb}
we have plotted the oscillation of the curvaton field about the
minimum of the potential for the quadratic  and the washboard
potential cases, for the same initial field value given by
$\sigma_*/M_p=0.1$. We can see that the
amplitude of oscillation in the washboard case
decreases faster than the quadratic case. Moreover, the frequency of
oscillation for the washboard curvaton is time dependent, it oscilates
about the constant frequency of the quadratic case.

\begin{figure}[h]
\begin{center}
\includegraphics[height=9.5cm,width=7.5cm]{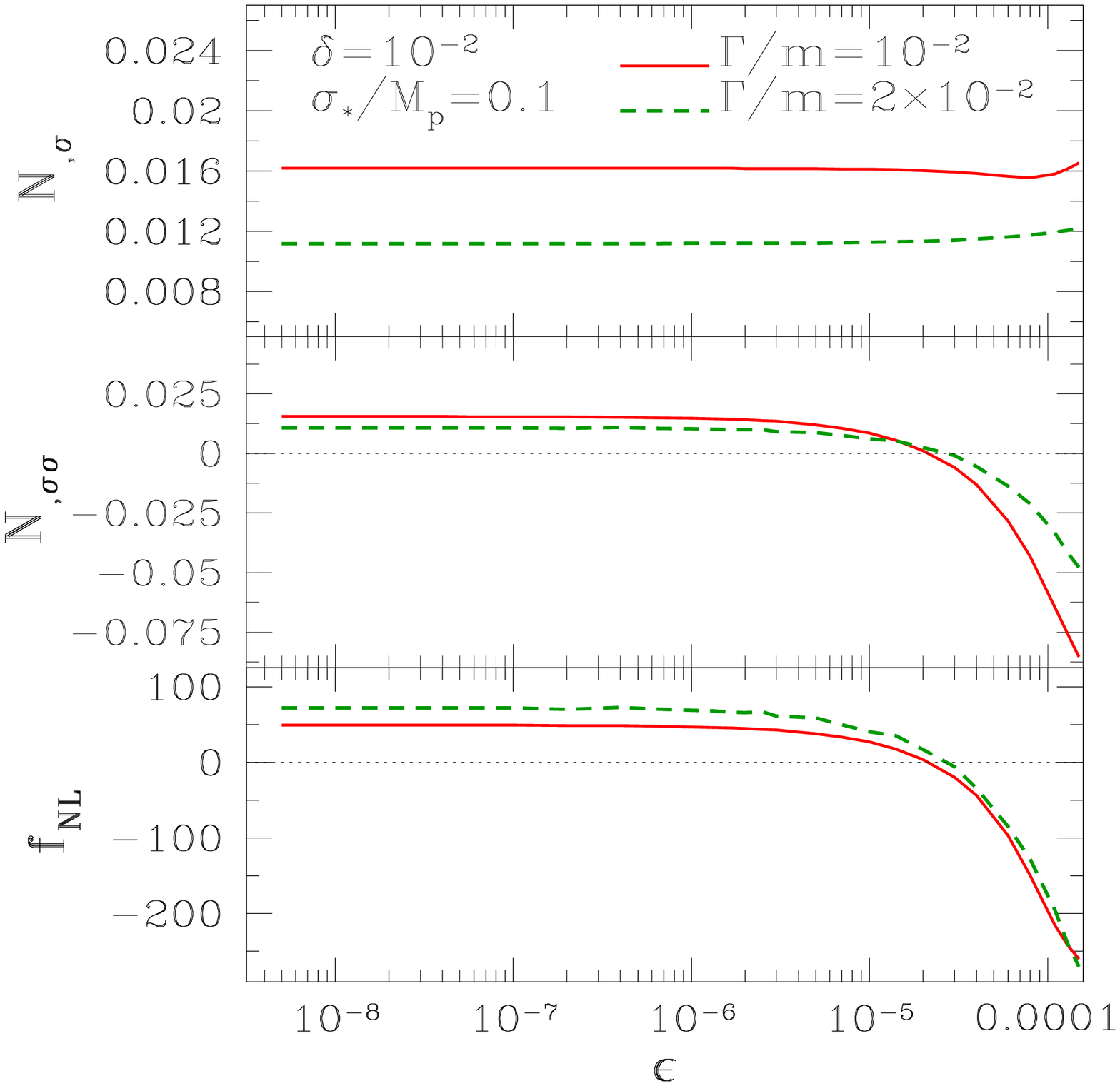}
\includegraphics[height=9.5cm,width=7.5cm]{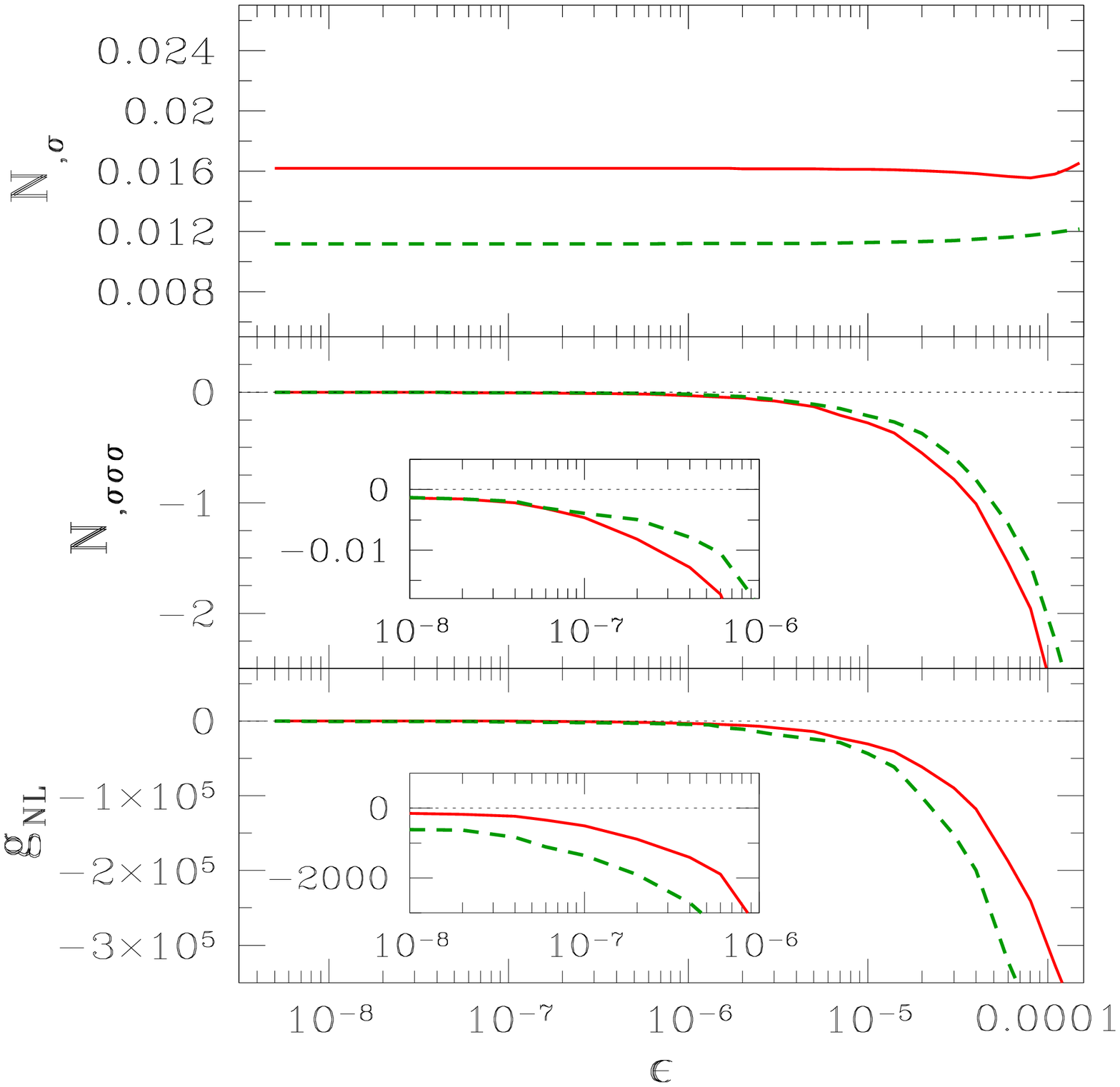}
\end{center}
\caption{$N_{,\sigma},\ N_{,\sigma\sigma} ,\
  N_{,\sigma\sigma\sigma}$, $f_{NL}$ and $g_{NL}$ are shown as
  functions of
  $\epsilon$ for the washboard potential, for fixed values of
  $\delta,\ \Gamma/m$ and  $\sigma_*/M_p$. We have shown plots for two
  different values of  $\Gamma/m$ in order to demonstrate the
  systematic variation as we change $\Gamma/m$.
}
\label{fig:varygamma}
\end{figure}

We now illustrate how the small features in the
curvaton potential play an important role for the
non-Gaussianity parameters. First we solve for $N_{,\sigma}$,
$N_{,\sigma\sigma}$ and
$N_{,\sigma\sigma\sigma}$ and then obtain $f_{NL}$ and $g_{NL}$ from them.
In general, we need to scan four independent parameters,
namely, $\Gamma/m, \ \sigma_*/M_p, \ \epsilon$ and $\delta$ in order to
satisfy observational constraints such as amplitude of perturbations
and the limits on $f_{NL}$ and $g_{NL}$. Our
strategy here is to fix $\Gamma/m, \ \sigma_*/M_p$ and $\delta$ and obtain
$f_{NL}$ and $g_{NL}$ as functions of $\epsilon$. Our results are
obtained for two values of $\Gamma/m$ and $\delta$
each, to understand how these parameters systematically affect
$f_{NL}$ and $g_{NL}$.

In the case of the quadratic potential $\Gamma/m$ is typically
required to be of the order of $10^{-8}$ for the amplitude of
perturbations to be  COBE normalized.
Evolving the equations numerically till the energy density of the
curvaton decreases to such small value is prohibitively time
consuming. Moreover, for
the purpose of capturing the essential features of the dependence of
$f_{NL}$ and $g_{NL}$ on $\epsilon$ and $\delta$, it is enough to fix
$\Gamma/m$ at a relatively larger value. To demonstrate this point, in
Fig.~{\ref{fig:varygamma}} we have plotted for $\Gamma/m = 10^{-2}$
and $2\times 10^{-2}$, how $N_{\sigma},\ N_{\sigma\sigma},\
N_{\sigma\sigma\sigma}$, $f_{NL}$
and $g_{NL}$ vary as functions of  $\epsilon$,  for fixed values of
$\delta=10^{-2}$ and $\sigma_*/M_p=0.1$. We can see that $\Gamma/m$
systematically changes the amplitudes of $f_{NL}$ and $g_{NL}$, but
does not alter the essential functional shapes. The correctness of the
numerical calculations are tested by ensuring that in the limit
$\epsilon \rightarrow 0$, $f_{NL}$ and $g_{NL}$ tend to their
analytically expected values for the quadratic potential, as clearly
seen in the figure.  As $\epsilon$ increases,
$f_{NL}$ and $g_{NL}$ get strongly affected and deviates from their
expectation from quadratic potential. $f_{NL}$ crosses over from
positive to increasingly negative values as $\epsilon$ increases.
On the other hand, $g_{NL}$ remains negative throughout but its
magnitude becomes  very large as $\epsilon$ increases.
The inset figures in the panels showing $N_{\sigma\sigma\sigma}$ and
$g_{NL}$ zoom in on the $\epsilon \rightarrow 0$ region to show them
approaching the negative values expected from quadratic potential.

\begin{figure}[h]
\begin{center}
\includegraphics[height=9.5cm,width=7.5cm]{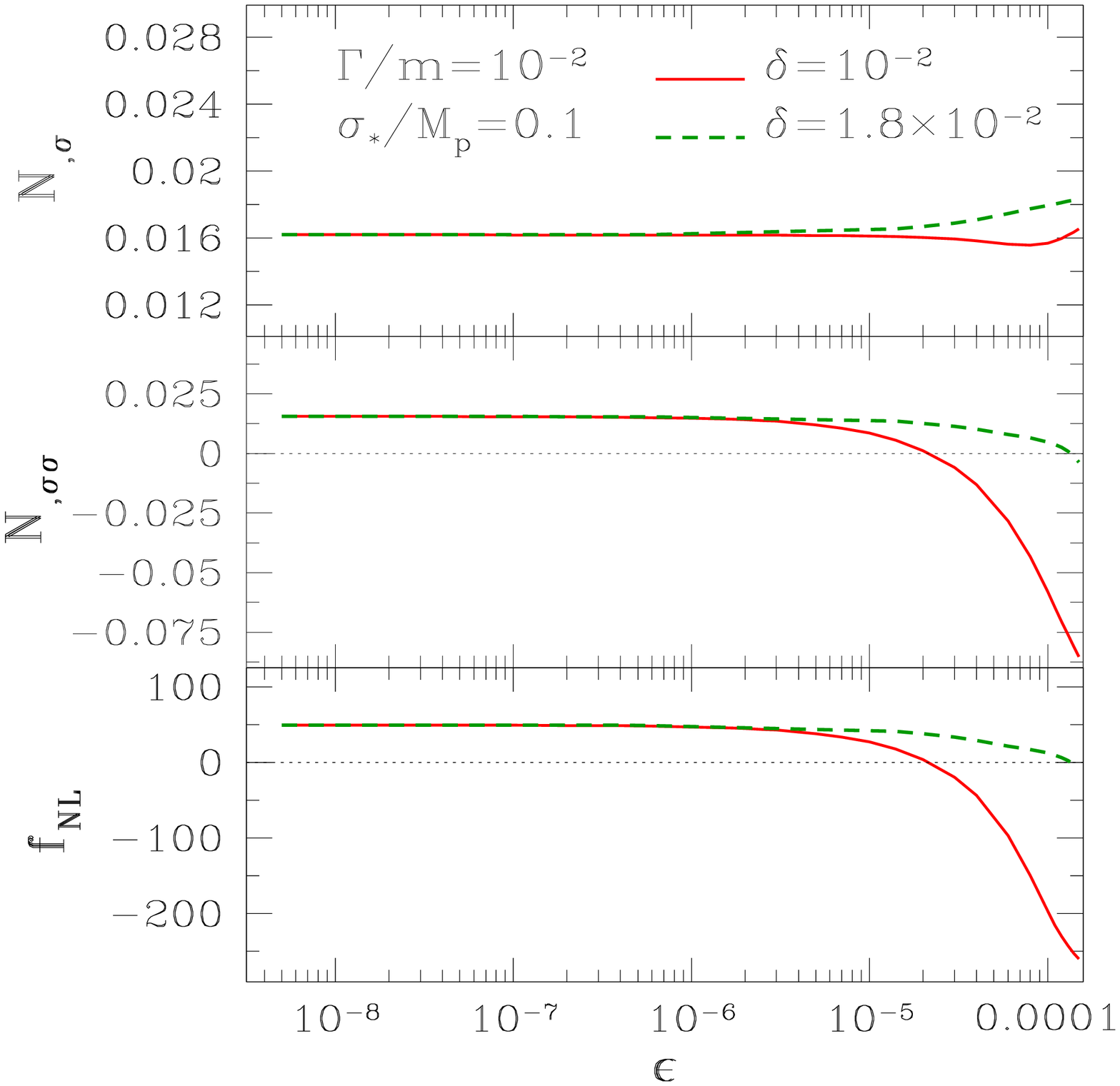}
\includegraphics[height=9.5cm,width=7.5cm]{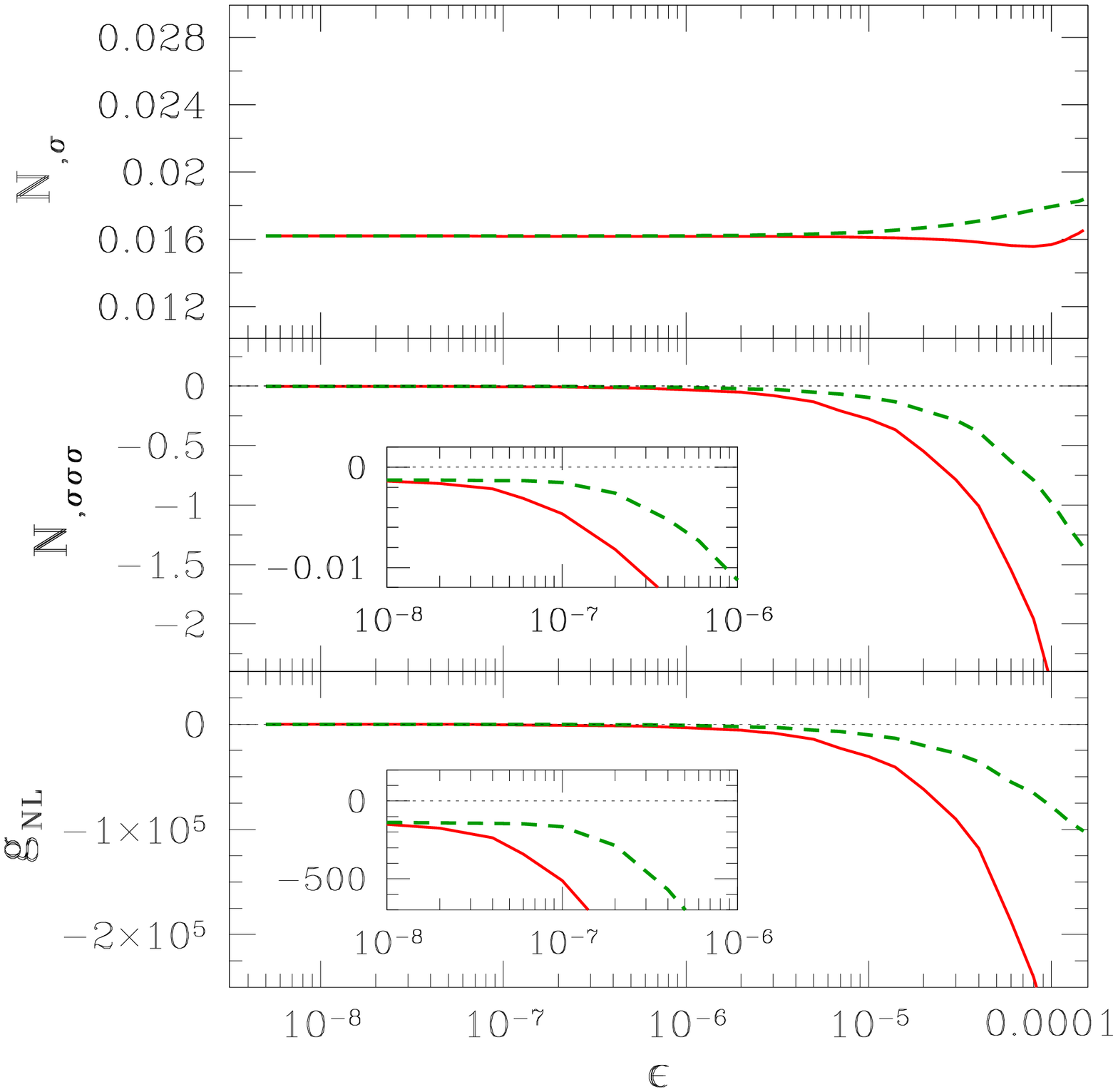}
\end{center}
\caption{Same as Fig. 2 but for two different
  values of $\delta$, with $\Gamma/m$ and $\sigma_*/M_p$ kept fixed. }
\label{fig:varydelta}
\end{figure}

Next, in Fig.~\ref{fig:varydelta} we have plotted $N_{\sigma},\
N_{\sigma\sigma},\ N_{\sigma\sigma\sigma}$, $f_{NL}$ and $g_{NL}$ as
functions of $\epsilon$, for two different values of $\delta$. We have
chosen $\delta=10^{-2}$ and $1.8\times 10^{-2}$ and
fixed $\Gamma/m = 10^{-2}$ and $\sigma_*/M_p=0.1$.
We see that for very small $\epsilon$, varying $\delta$ has little
effect on the behavior of $f_{NL}$ and $g_{NL}$. This can be explained
by the fact that $\epsilon \rightarrow 0$ kills off the oscillations
superimposed on the potential, regardless of the frequency of
oscillations which is controlled by $\delta$. At relatively larger
values of $\epsilon$, the effect of $\delta$ becomes prominent. As
the deviation of $f_{NL}$ and $g_{NL}$ from the quadratic potential
behavior increases as $\delta$ decreases, due to the increase in the
frequency of the oscillations in the potential. As in
Fig.~\ref{fig:varygamma} the inset figures in the panels showing
$N_{\sigma\sigma\sigma}$ and $g_{NL}$ zoom in on the $\epsilon
\rightarrow 0$ region to show them approaching the negative values
expected from quadratic potential.

%%%%%%%%%%%%%%%%%%%%%%%%%%%%%%%%%%%%%%%%%%%%%%%%%%%%%%%%%%%%%%%%%%%%
\subsection{Single-feature curvaton model}

In this subsection we consider a curvaton model with the potential
given by
\e
V(\sigma)=\half m^2 \sigma^2 \(1+{c\over 1+(\sigma/M)^{2n}}\),
\q
where $n>0$, and $M$ is an energy
scale which measures the position of the feature. In the regime
$\sigma \gg M$ or $\sigma\ll M$, the curvaton potential has
a quadratic form, but around $\sigma\sim M$ the potential deviates
from quadratic form. As in the previous subsection, we define a
reduced potential, as follows
\e
V({\tilde \sigma})=\half {\tilde \sigma}^2\(1+{c\over 1+({\tilde
\sigma}/d)^{2n}}\),
\label{eqn:psb}
\q
where
\m
d={M\over \sigma_*}.
\n
\begin{figure}[h]
\begin{center}
\includegraphics[height=6.5cm,width=6.cm]{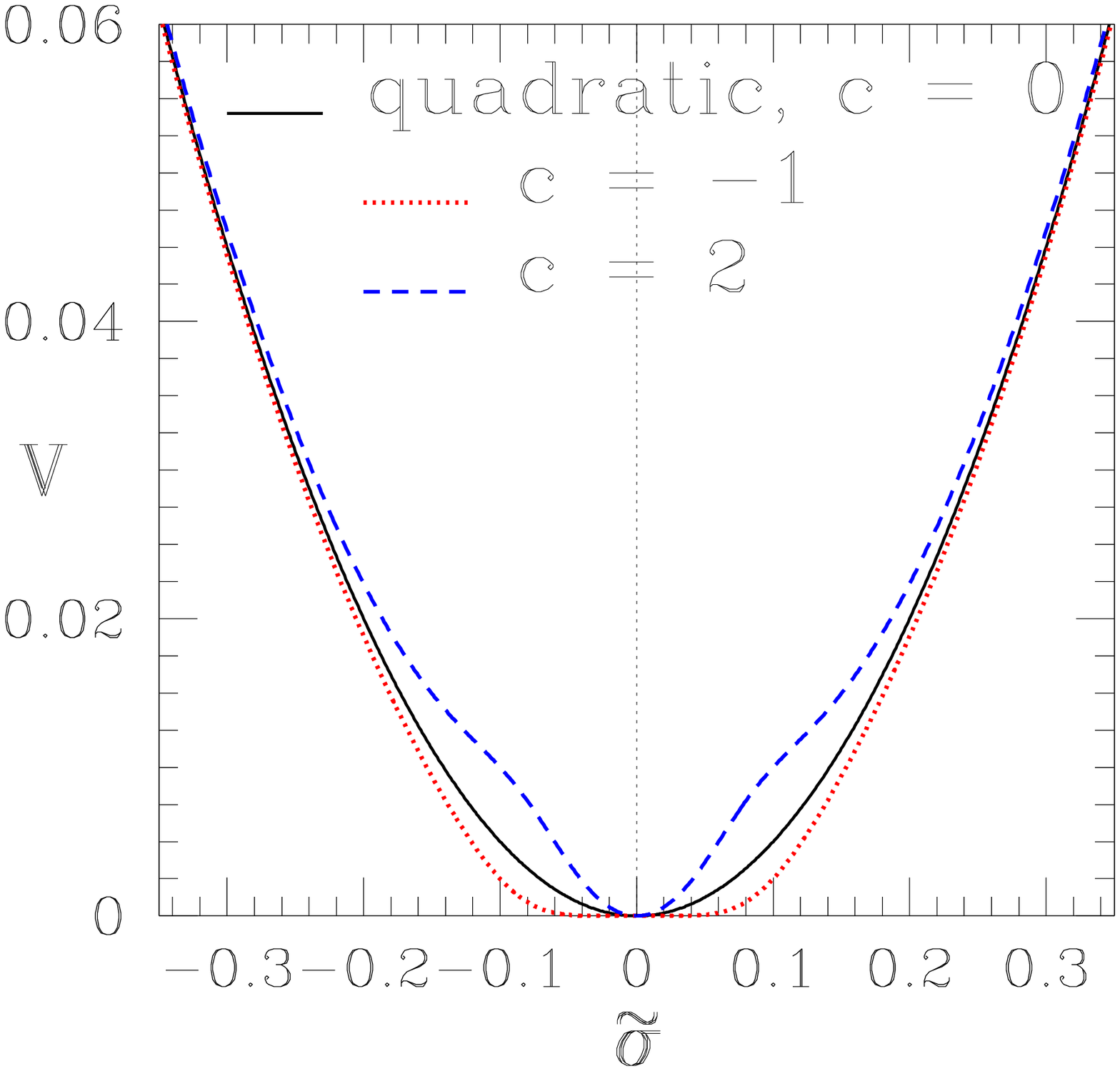}
\includegraphics[height=6.5cm,width=9.cm]{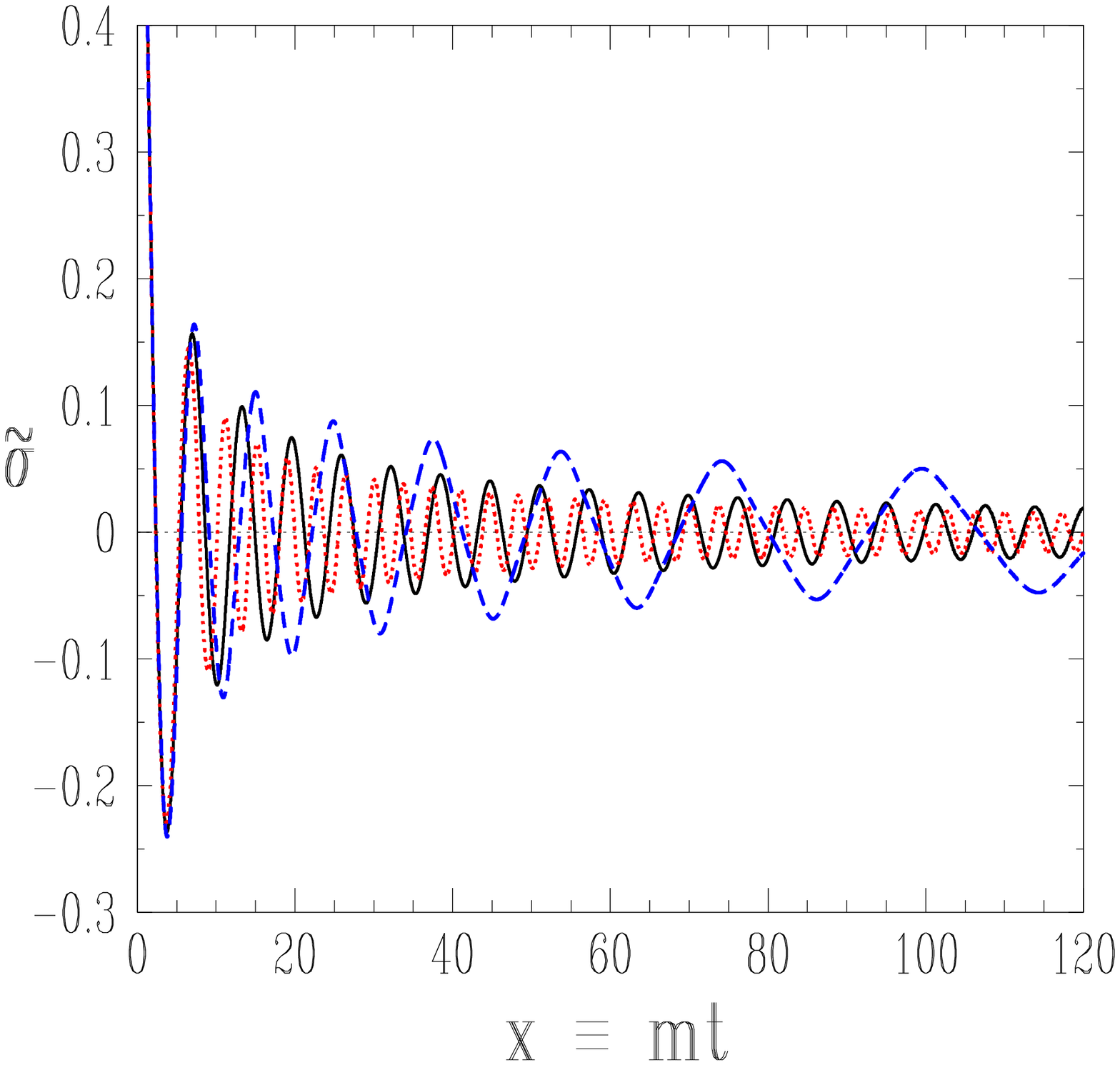}
\end{center}
\caption{
The single feature curvaton potential given by
  Eq.~(3.7) is shown on the left panel for comparision with the
  corresponding quadratic one. $d$ is fixed to be $0.1$ and we have
  plotted for $c=2$ and $-1$ to show how the nature of the
  feature changes with the sign of $c$. The right panel shows the
  corresponding curvaton oscillations about the potential minimum,
  with the same initial field value given by $\sigma_*/M_p=0.1$.
}
\label{fig:psb}
\end{figure}
The reduced potential given by Eq.~(\ref{eqn:psb}) is shown in
Fig.~\ref{fig:psb} for $n=2$ and $d=0.1$.  The nature of the feature
depends on the sign of $c$. If $c$ is positive, then there is a bump,
whereas, a negative $c$ changes the slope of the potential to make it
flatter around some scale set by the parameter $d$.
%${\tilde \sigma}=0$.

The equation of motion for the reduced curvaton field becomes
\m
{\tilde \sigma}''&+&3N'{\tilde \sigma}'+\(1+c{1-(n-1)({\tilde
    \sigma}/d)^{2n}  \over (1+({\tilde \sigma}/d)^{2n})^2} \) {\tilde
  \sigma}=0.
\n
We restrict our analysis here to $n=2$. If $d\ll 1$ and the initial
curvaton field value is large enough, then the curvaton evolves
linearly prior to its oscillation. We choose $d=0.1$. As in the
washboard curvaton model, we choose $\Gamma_\sigma/m$ to be
$10^{-2}$ and $\sigma_*/M_p=10^{-1}$. Then we solve for $f_{NL}$,
scanning the parameter $c$.
\begin{figure}[h]
\begin{center}
\includegraphics[height=9.5cm,width=7.5cm]{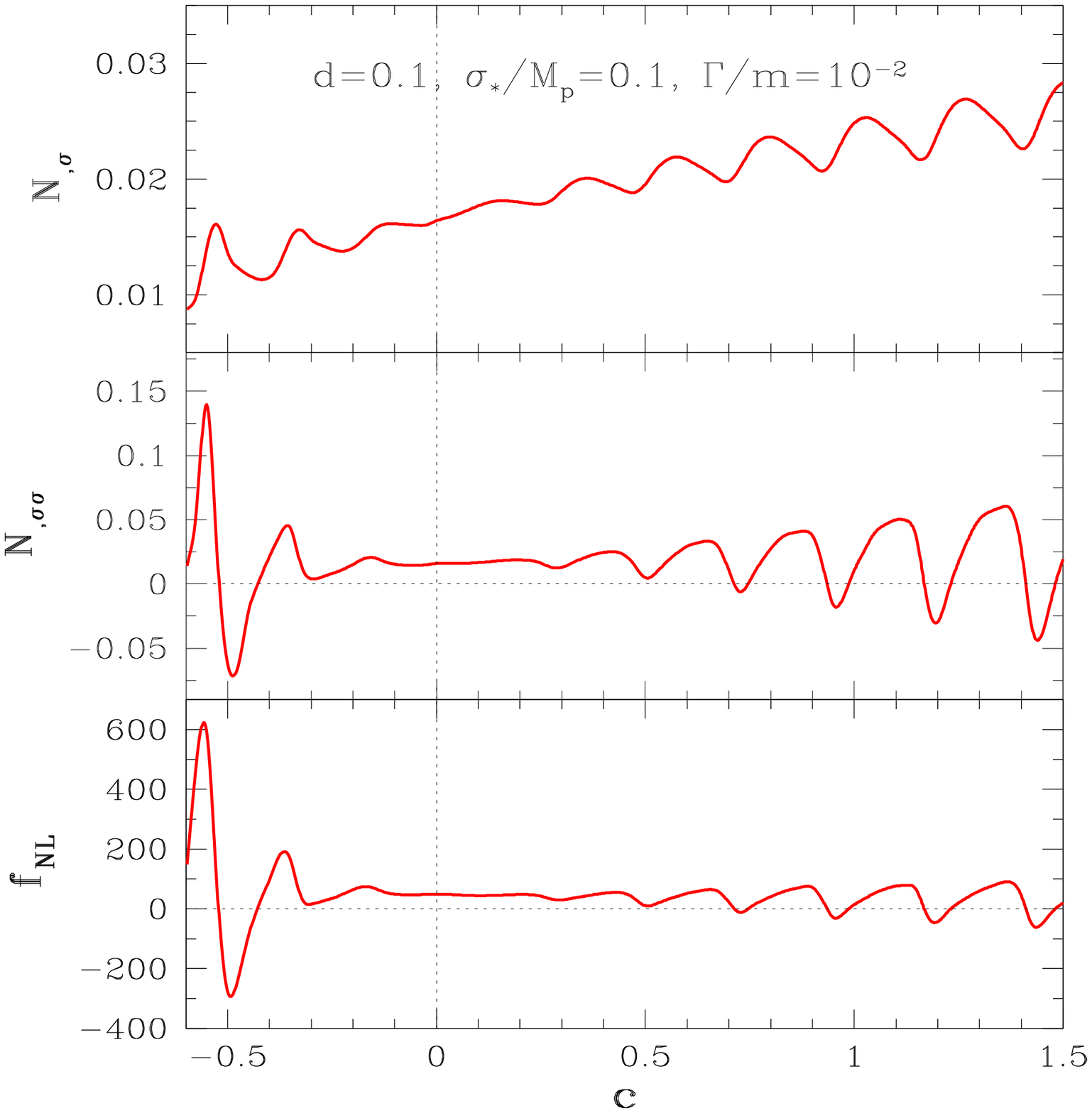}
\includegraphics[height=9.5cm,width=7.5cm]{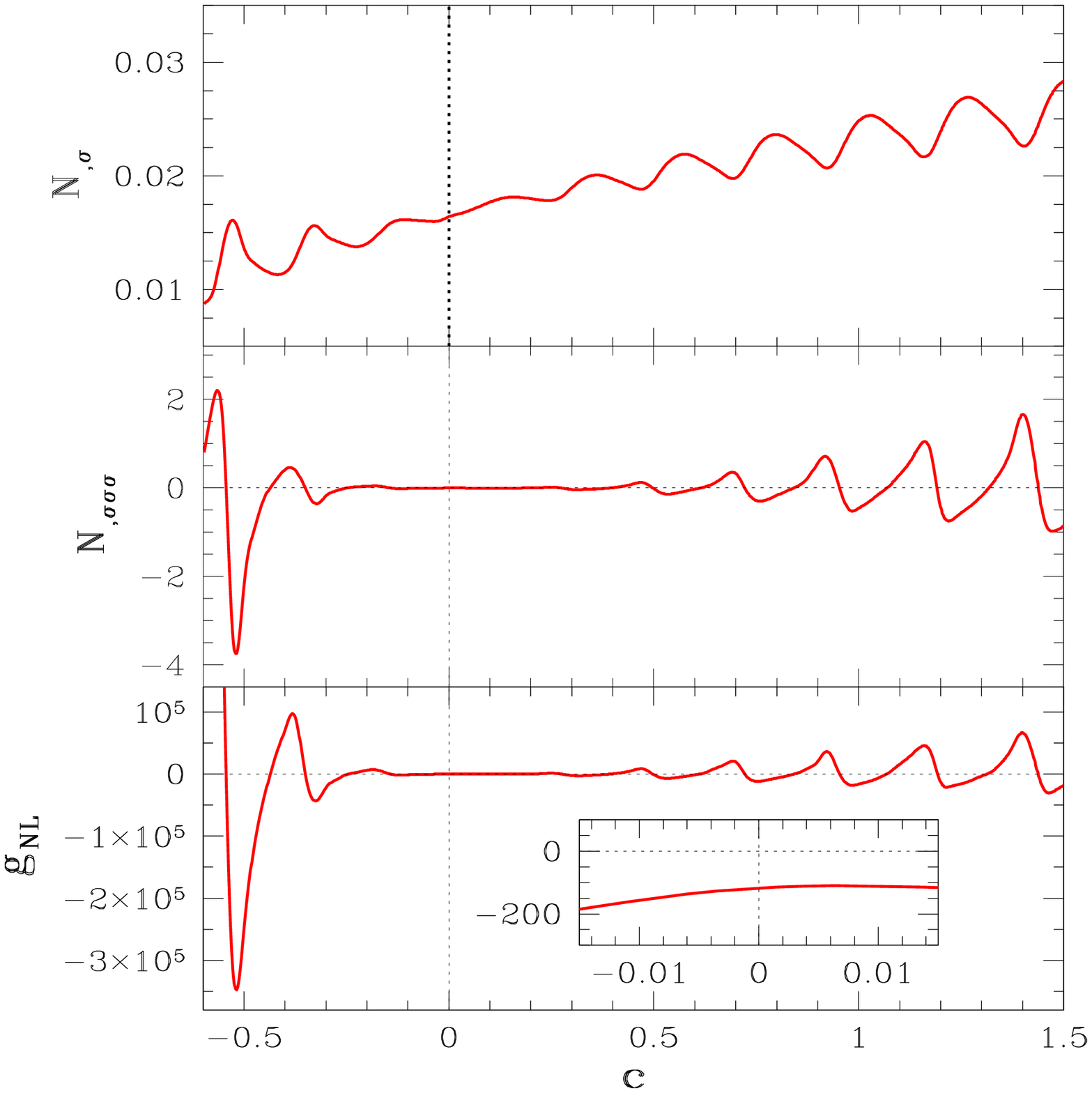}
\end{center}
\caption{
$N_{,\sigma},\ N_{,\sigma\sigma} ,\
  N_{,\sigma\sigma\sigma}$, $f_{NL}$ and $g_{NL}$ are shown as
  functions of $c$ for the single-feature potential, with  $d,\
  \Gamma/m$ and   $\sigma_*/M_p$ kept fixed.
}
\label{fig:sovv}
\end{figure}
On the left panel of Fig. \ref{fig:sovv}, we have plotted $N_{,\sigma}$,
$N_{,\sigma\sigma}$ and $f_{NL}$ for the single-feature
potential. As shown in the figure, $f_{NL}$ oscillates about zero with
increasing amplitude as $\abs{c}$ increases.
We can see $f_{NL}$ flattening out and approaching the expected value
from the quadratic potential as $\abs{c}\rightarrow 0$.
On the right panel of the same fugure we have plotted  $N_{,\sigma}$,
$N_{,\sigma\sigma\sigma}$ and $g_{NL}$. We again obtain oscillatory
behavior of
$g_{NL}$ as $c$ varies. Similar to $f_{NL}$, we can see the curve
flattening out near $c=0$ for  $g_{NL}$ to assumes the value expected
from quadratic potential, as shown in the inset figure on the right
bottom panel.

%%%%%%%%%%%%%%%%%%%%%%%%%%%%%%%%%%%%%%%%%%%%%%%%%%%%%%%%%%%%%%%%%%%%%%%%%%
\section{Conclusion and discussion}

We have studied two new curvaton models in this paper.
The first is the washboard model where the potential has tiny
oscillations superimposed on the quadratic form, and the second one
has a potential with two quadratic regimes having different mass
scales separated by either a bump or a flattening of the potential.
For the washboard model we have investigated in detail how the two
parameters that control the
oscillations, namely, the amplitude and the frequency, affect the
non-linear corrections to the curvature perturbation via their effect
on $f_{NL}$ and $g_{NL}$.  We have shown
that the relation $g_{NL} \propto -f_{NL}$, which holds for the
quadratic potential, is no longer valid in this case. We also found
that there is a wide range of both positive and negative values for
$f_{NL}$, while $g_{NL}$ remains negative but its magnitude can be
very large depending on the model parameters. In
comparision, the quadratic potential
restricts $f_{NL}$ to be positive and $g_{NL}$ to be negative.
For the single-feature model we have again calculated $f_{NL}$ and
$g_{NL}$, and demonstrated that they strongly depend on the strength
of the feature and oscillate as the strength increases.

What is new in the models considered here is that the curvation
motion as it oscillates about the potential minimum is non-linear,
unlike other models that have been considered so far in the
literature. The results that we have found have interesting
implications for searches for non-Gaussianity in observational data.
The fact that  $f_{NL}$ can switch sign at some parameter values
implies that it is possible that the non-linear contributions to the
curvature perturbation could be coming from $g_{NL}$ alone, with
$f_{NL}$ being close to zero. Similar result was obtained
in~\cite{Enqvist:2005pg} in the context of curvaton potential with
non-linear corrections to the quadratic term. It is also possible
that both $f_{NL}$ and $g_{NL}$ contribute comparably with the same or
opposite signs.  The present work thus throws up the need to
understand different sources of primordial non-Gaussianity and how
they can be distinguished in the observational data. It is important
to devise observables which can distinguish them. Such studies have
been initiated in~\cite{Chingangbam:2009vi, Matsubara:2010te}. 
We also want to mention that $f_{NL}$ and $g_{NL}$ are controlled by
two independent geometric quantities which characterize the
hyper-surface in field space on which multi-field inflation ends. This
has the implication that all of the possible results for $f_{NL}$ and
$g_{NL}$ in curvaton
models can be realized by tuning these two independent geometric
quantities, as shown in~\cite{Huang:2009vk,Sasaki:2008uc,Huang:2009xa}. 

In principle, the three free parameters in the washboard model,
$\Gamma/m, \epsilon$ and $\delta$, can be
constrained by using the observational constraints on $f_{NL}$,
$g_{NL}$ and the amplitude of perturbations. The parameters $c$ and
$d$ in the single-feature model can be similarly constrained. However,
such a full scan of
the parameter space is beyond the scope of the present analysis. Our
purpose in this paper has been to understand the systematic behaviors of
$f_{NL}$ and $g_{NL}$ as functions of the model parameters. We will
tackle the problem of scanning the parameter space in a future work.

%%%%%%%%%%%%%%%%%%%%%%%%%%%%%%%%%%%%%%%%%%%%%%%%%%%%%%%%%%%%%%%%%%%%%%%%%%
\vspace{1.cm}

\noindent {\bf Acknowledgments}

\vspace{.2cm}

P.C. is supported by the National Research Foundation of Korea(NRF)
grant funded by the Korea government(MEST) (No. 2009-0062868).
QGH is supported by the project of Knowledge Innovation Program of
Chinese Academy of Science. The numerical calculation in this work
was carried out on the QUEST cluster computing facility at Korea
Institute for Advanced Study.

%%%%%%%%%%%%%%%%%%%%%%%%%%%%%%%%%%%%%%%%%%%%%%%%%%%%%%%%%%%%%%%%%%%%%%%%%%
\vspace{1.cm}

\appendix

\section{Curvaton model with quadratic potential}
%\label{ap}

For the curvaton model with quadratic potential, from
\cite{Lyth:2005fi}, we have
\m
f_{NL}&=& {5\over 4f_D}-{5\over 3}-{5\over 6}f_D,\\
g_{NL}&=& -{25\over 6f_D}+{25\over 108}+{125\over 27}f_D+{25\over 18}f_D^2,
\n
where
\e
f_D={3\Omega_{\sigma,D}\over 4-\Omega_{\sigma,D}}.
\q
After inflation the universe is dominated by radiation and
the Hubble parameter is related to the cosmic time $t$ by
$H={1\over 2t}$. The equation of motion of curvaton field
with quadratic potential becomes \e {\tilde
\sigma}''+{3\over 2x}{\tilde \sigma}'+{\tilde \sigma}=0, \q
whose solution is \e {\tilde \sigma}=2^{1/4}\Gamma(5/4)
x^{-1/4} J_{1/4}(x), \q where $J_{\nu}(x)$ is the Bessel
function of the first kind. Therefore the energy density of
curvaton is given by \e \rho_\sigma=\half m^2\sigma_*^2
\({\tilde \sigma}^2+({d{\tilde \sigma}\over dx})^2\)
={\Gamma^2(5/4)\over \sqrt{2}} m^2\sigma_*^2
x^{-1/2}\(J_{1/4}^2(x)+J_{5/4}^2(x)\). \q Adopting the
sudden decay approximation, we have \e
\Omega_{\sigma,D}={\rho_\sigma(x_D)\over 3M_p^2\Gamma_\sigma^2}
\simeq 0.35{\sigma_*^2\over M_p^2}\sqrt{m\over
\Gamma_\sigma}, \q in the limit of $x_D=\half {m\over
\Gamma_\sigma}\gg 1$. In the literatures,
$\Omega_{\sigma,D}={\sigma_*^2\over 6M_p^2}\sqrt{m\over
\Gamma_\sigma}$ which is roughly half of our exact retult.

%%%%%%%%%%%%%%%%%%%%%%%%%%%%%%%%%%%%%%%%%%%%%%%%%%%%%%%%%%%%%%%%%%%%%%%%%%%%

\end{document}